\documentclass[conference]{IEEEtran}
\IEEEoverridecommandlockouts
\usepackage{amsmath,amssymb,amsfonts}
\usepackage{algorithmic}
\usepackage{caption}
\usepackage{graphicx}
\usepackage{textcomp}
\usepackage{xcolor}
\usepackage{hyperref}
\usepackage{float}
\usepackage{tikz}
\usetikzlibrary{svg.path}
\usepackage{scalerel}

\definecolor{orcidlogocol}{HTML}{A6CE39}
\tikzset{
  orcidlogo/.pic={
    \fill[orcidlogocol] svg{M256,128c0,70.7-57.3,128-128,128C57.3,256,0,198.7,0,128C0,57.3,57.3,0,128,0C198.7,0,256,57.3,256,128z};
    \fill[white] svg{M86.3,186.2H70.9V79.1h15.4v48.4V186.2z}
                 svg{M108.9,79.1h41.6c39.6,0,57,28.3,57,53.6c0,27.5-21.5,53.6-56.8,53.6h-41.8V79.1z M124.3,172.4h24.5c34.9,0,42.9-26.5,42.9-39.7c0-21.5-13.7-39.7-43.7-39.7h-23.7V172.4z}
                 svg{M88.7,56.8c0,5.5-4.5,10.1-10.1,10.1c-5.6,0-10.1-4.6-10.1-10.1c0-5.6,4.5-10.1,10.1-10.1C84.2,46.7,88.7,51.3,88.7,56.8z};
  }
}
\newcommand\orcidicon[1]{\href{https://orcid.org/#1}{\mbox{\scalerel*{
\begin{tikzpicture}[yscale=-1,transform shape]
\pic{orcidlogo};
\end{tikzpicture}
}{|}}}}
\def\BibTeX{{\rm B\kern-.05em{\sc i\kern-.025em b}\kern-.08em
    T\kern-.1667em\lower.7ex\hbox{E}\kern-.125emX}}

\begin{document}
\title{Machine Learning-Driven User Localization in RIS-Assisted Wireless Systems}

\author{
    \IEEEauthorblockN{
        Md Tarek Hassan \orcidicon{0000-0002-9719-0204}, 
        Dmitry Zelenchuk \orcidicon{0000-0003-2866-629}, and
        Muhammad Ali Babar Abbasi \orcidicon{0000-0002-1283-4614}
    }
    \IEEEauthorblockA{
        Centre for Wireless Innovation (CWI), School of Electronics, Electrical Engineering and Computer Science (EEECS), \\Queen’s University Belfast, U.K.\\
        Emails: \{mhassan15, d.zelenchuk, m.abbasi\}@qub.ac.uk
    }
}


\maketitle

\begin{abstract}
The sixth generation (6G) targets ultra-reliable, low-latency (URLLC) gigabit connectivity in mmWave bands, where fragile directional channels demand precise beam alignment. Reconfigurable Intelligent Surfaces (RIS) offer a promising solution by reshaping wave propagation and extending coverage, yet they enlarge the beam-search space at the base station (BS), making exhaustive sweeps inefficient due to control overhead and latency. This challenge motivates machine learning (ML)-driven user localization using minimal probing beams. This paper presents an ML-based localization framework for RIS-assisted communication at 27 GHz. A $20\times20$ RIS reflects signals from a core network connected BS and sweeps beams across the 0–90° elevation plane, divided into four angular sectors. A dataset is constructed by recording received signal power, $P_r$ (dBm), across user locations and trained with multiple regressors, including Decision Tree (DT), Support Vector Regressor (SVR), K-Nearest Neighbor (KNN), XGBoost, Gradient Boosting, and Random Forest. In operation, an unknown user in the same plane measures four received power values (one per sector) and reports them to the pre-trained RIS controller, which predicts the user’s angular position in real time. Extensive evaluation using mean absolute error (MAE), root mean squared error (RMSE), and $R^2$ score demonstrates the framework’s accuracy. The DT model achieves an MAE of 4.8° with $R^2$ of 0.96, while other models also deliver competitive performance, 70\%-86\%. Predicted radiation patterns, including main-lobe alignment between (52–55)°, closely track ground truth, validating the estimation accuracy. The proposed framework reduces beam probing requirements, enables faster alignment, and achieves lower latency for RIS-assisted 6G networks.
\end{abstract}

\section{Introduction}
Reconfigurable intelligent surfaces (RISs) have emerged as a key enabler for the sixth generation (6G) systems by shaping the radio environment through programmable, low-power metasurfaces that impose controllable phase (and, in practice, amplitude) profiles on impinging waves \cite{9140329}. Beyond coverage extension and blockage mitigation, RISs open new possibilities for high-accuracy radio localization by generating additional, well-structured non-line-of-sight (NLOS) paths and by concentrating energy along designed directions \cite{umer_ris_localization_2024}. Concurrently, machine learning (ML) has matured as a powerful paradigm for RF-based positioning, offering data-driven inference from features such as received signal strength, time/angle of arrival, and channel state information (CSI), while adapting to hardware imperfections and environment dynamics \cite{burghal_ml_loc_survey_2020}. Integrating ML with RIS control has emerged as a promising direction for enhancing user localization accuracy in next-generation wireless networks. Instead of relying solely on geometric or signal-model-based estimation, ML enables data-driven learning of the complex mapping between received signal features and user positions, even under dynamic and NLOS conditions \cite{3gppTR38843}. By leveraging channel measurements, reflected power patterns, and spatial correlations, ML models can infer user locations with reduced probing overhead and higher robustness to environmental uncertainty. Within RIS-assisted systems, ML further supports adaptive beam management and intelligent reflection configuration, allowing the network to jointly optimize sensing and communication \cite{Giordani2019BeamTutorial}.
Recent works demonstrate ML-aided RIS localization in challenging NLOS and near-field regimes, leveraging probabilistic filters, sparse recovery, and learned optimizers to achieve sub-meter and in simulations even sub-centimeter accuracy \cite{9625826, 9838395}.  This work develops an ML-driven user-localization framework for RIS-assisted links and validates it on a 27\,GHz testbed with a $20 \times 20$ RIS. The main contributions of this paper are summarized as:
\begin{itemize}
  \item We present a low-overhead RIS probing policy that sweeps a four-sector codebook over $0^\circ\!-\!90^\circ$, extracts a per-sector received-power feature vector $\mathbf{P}_r$, and performs ML inference using regression models.
  \item We show the ML models' performance and analyze the radiation pattern from the predicted user location, comparing the actual location of the user with minimal beam probing.  
\end{itemize}
\begin{figure}[!htbp]
\centering
\includegraphics[width=0.45\textwidth]{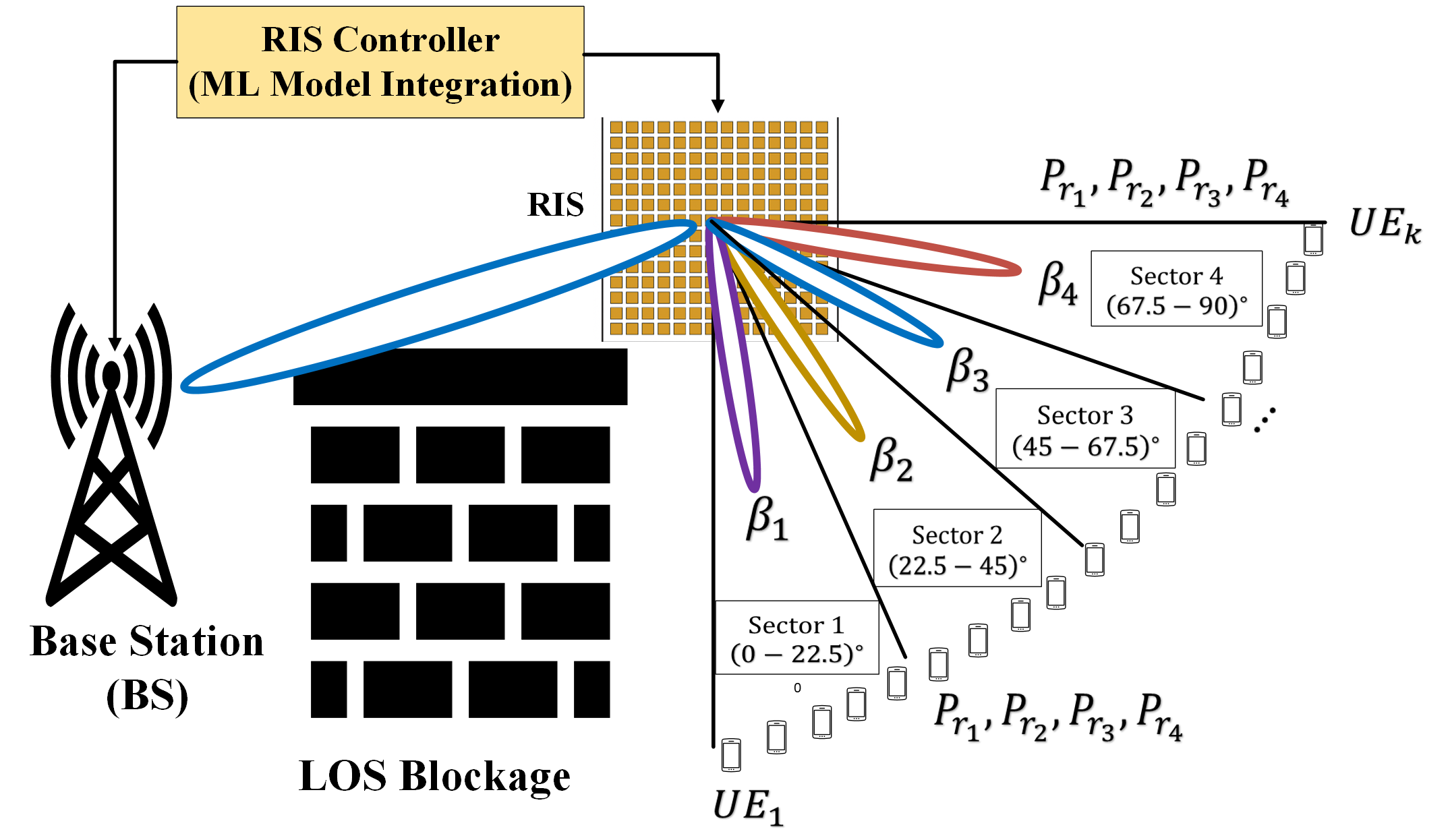}
\caption{RIS-aided wireless communication system model.}
\label{fig:Scenario}
\vspace{-5 pt}
\end{figure}
\section{System Model}
\noindent
Fig.~\ref{fig:Scenario} illustrates a RIS-assisted wireless communication system integrated with ML based RIS controller. The Base Station (BS) communicates with users located beyond a LOS blockage. To overcome this obstruction, the $20 \times 20$ RIS is deployed on a nearby structure to reflect and steer the BS signals toward the desired users. The RIS controller dynamically adjusts the phase shifts of the reflecting elements to form four directional beams, each covering a specific angular sectors. These beams ensure spatial diversity and maintain reliable connections even under blockage scenarios. The controller, aided by the ML model, estimates the user location as elevation and azimuth angles from power vectors corresponding to the likelihood of user presence or optimal signal path in each sector. By adaptively selecting the beam with the highest probability, the system enhances link reliability, signal quality, and overall communication performance in NLOS environments.
\begin{table}[!htbp]
\centering
\caption{Simulation Parameters}
\begin{tabular}{ll}
\hline
\textbf{Parameters Used} & \textbf{Value} \\
\hline
$G_{t}$ (Tx Gain) and $G_{r}$ (Rx Gain) & $15$ dBi \\
$M, N$ (RIS Elements) & $20, 20$ \\
Element spacing $d_{x}$ and $d_{y}$ & $4.6$ mm \\
Carrier frequency $f$ & $27$ GHz \\
$P_{t}$ (Tx Power) & $10$ dBm \\
Reflection coefficient amplitude $\Gamma_{\text{amp}}$ & $0.7$ \\
$d_{1}$ (Tx--RIS distance) and $d_{2}$ (RIS--Rx distance) & $2.3$ m \\
$\theta_{t}$ (Tx elevation) and $\varphi_{t}$ (Tx azimuth) & $0^{\circ}$ \\
$\theta_{r}$ (Rx azimuth) & $0^{\circ}-90^{\circ}$ \\
$\varphi_{r}$ (Rx azimuth) & $180^{\circ}$ \\
\hline
\end{tabular}
\label{tab:sim_params}
\end{table}
\section{Results and Discussion}
We run simulations in Python and MATLAB using the setup in Table~\ref{tab:sim_params}. 
Table~\ref{tab:model_comparison} summarizes the performance of random forest (RF), decision tree (DT), K-nearest neighbors (KNN), support vector regressor (SVR), XGBoost (XGB), and gradient boosting (GB) using three metrics: 
mean absolute error (MAE), average angular deviation in degrees, root mean squared error (RMSE), averages squared errors and thus penalizes large mistakes, and $R^2$ is explained variance, with $1$ indicating a perfect fit. 
DT performs best, achieving $\mathrm{MAE}\approx 4.8^\circ$ and $R^2\approx 0.96$.  The remaining models yield $\mathrm{MAE}$ between $7.5^\circ$ and $12.5^\circ$ and $R^2$ between $0.70$ and $0.87$. RMSE follows the same ordering as MAE but highlights outliers more strongly, reinforcing the robustness of the DT model.

\begin{table}[!htbp]
\centering
\caption{Performance Comparison of ML Models}
\begin{tabular}{lccc}
\hline
\textbf{Model} & \textbf{MAE ($^{\circ}$)} & \textbf{RMSE} & \textbf{$R^{2}$} \\
\hline
DT     & 4.800 & 6.066  & 0.959 \\
SVR               & 7.547 & 11.271 & 0.860 \\
KNN               & 8.054 & 12.509 & 0.828 \\
XGB           & 8.522 & 11.774 & 0.848 \\
GB & 10.390 & 14.197 & 0.779 \\
RF     & 12.458 & 16.461 & 0.702 \\
\hline
\end{tabular}
\label{tab:model_comparison}
\end{table}
\begin{figure}[!htbp]
\centering
\includegraphics[width=0.9\linewidth]{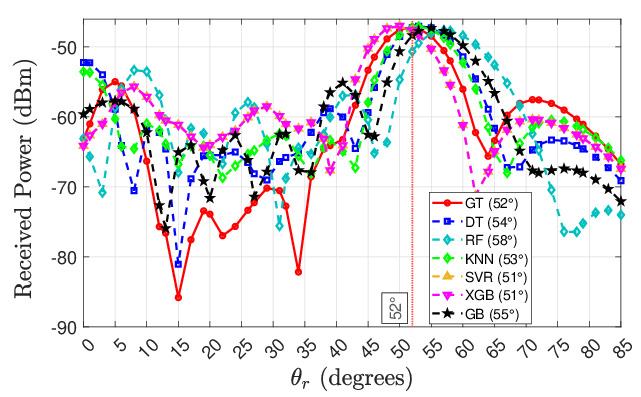}
\caption{Prediction comparison among classical ML models in terms of radiation pattern.}
\label{fig:prediction}
\end{figure}
Fig.~\ref{fig:prediction} compares the ground truth (GT) received power distribution with predictions from several ML models across the receiver angle $\hat\theta_r$. The GT shows a peak near $52^\circ$, while models such as GB and XGB accurately follow this trend with minor deviations. 
Predicted peak angles (from the legend) are: DT $54^\circ$ $(+2^\circ)$, RF $58^\circ$ $(+6^\circ)$, KNN $53^\circ$ $(+1^\circ)$, SVR $51^\circ$ $(-1^\circ)$, XGB $51^\circ$ $(-1^\circ)$, and GB $55^\circ$ $(+3^\circ)$. Across $45^\circ$–$65^\circ$, XGB, SVR, and KNN track the GT lobe closely (alignment error $\leq 1$–$2^\circ$), GB is slightly biased ($\approx 3^\circ$), while RF shows the largest shift and more ripple at lower angles ($0^\circ$–$30^\circ$). Overall, XGB, SVR, and KNN best preserve the GT peak location and shape; RF and DT exhibit higher angular bias and variability.

\section{Conclusion}
The proposed method achieves precise user localization in RIS-assisted systems using only four probing beams, significantly reducing beam-training overhead and alignment delay. Such minimal-probe, data-driven localization meets critical requirements for energy efficiency, responsiveness, and reliability in RIS-enabled 6G communication.
\bibliographystyle{IEEEtran}
\bibliography{Biblio}

\begin{thebibliography}{1}
\providecommand{\url}[1]{#1}
\csname url@samestyle\endcsname
\providecommand{\newblock}{\relax}
\providecommand{\bibinfo}[2]{#2}
\providecommand{\BIBentrySTDinterwordspacing}{\spaceskip=0pt\relax}
\providecommand{\BIBentryALTinterwordstretchfactor}{4}
\providecommand{\BIBentryALTinterwordspacing}{\spaceskip=\fontdimen2\font plus
\BIBentryALTinterwordstretchfactor\fontdimen3\font minus \fontdimen4\font\relax}
\providecommand{\BIBforeignlanguage}[2]{{%
\expandafter\ifx\csname l@#1\endcsname\relax
\typeout{** WARNING: IEEEtran.bst: No hyphenation pattern has been}%
\typeout{** loaded for the language `#1'. Using the pattern for}%
\typeout{** the default language instead.}%
\else
\language=\csname l@#1\endcsname
\fi
#2}}
\providecommand{\BIBdecl}{\relax}
\BIBdecl

\bibitem{9140329}
M.~Di~Renzo, A.~Zappone, M.~Debbah, M.-S. Alouini, C.~Yuen, J.~de~Rosny, and S.~Tretyakov, ``Smart radio environments empowered by reconfigurable intelligent surfaces: How it works, state of research, and the road ahead,'' \emph{IEEE Journal on Selected Areas in Communications}, vol.~38, no.~11, pp. 2450--2525, 2020.

\bibitem{umer_ris_localization_2024}
A.~Umer, I.~M{\"u}{\"u}rsepp, M.~M. Alam, and H.~Wymeersch, ``Role of reconfigurable intelligent surfaces in {6G} radio localization: Recent developments, opportunities, challenges, and applications,'' \emph{arXiv:2312.07288}, 2024.

\bibitem{burghal_ml_loc_survey_2020}
D.~Burghal, A.~T. Ravi, V.~Rao, A.~A. Alghafis, and A.~F. Molisch, ``A comprehensive survey of machine learning based localization with wireless signals,'' \emph{arXiv:2012.11171}, 2020.

\bibitem{3gppTR38843}
``{Study on {A}rtificial {I}ntelligence ({AI})/{M}achine Learning ({ML}) for {NR} air interface},'' 3rd Generation Partnership Project (3GPP), Technical Report 3GPP TR 38.843, 2024, technical Specification Group Radio Access Network.

\bibitem{Giordani2019BeamTutorial}
M.~Giordani, M.~Polese, A.~Roy, D.~Castor, and M.~Zorzi, ``A {T}utorial on {B}eam {M}anagement for {3GPP} {NR} at mm{W}ave {F}requencies,'' \emph{IEEE Communications Surveys \& Tutorials}, vol.~21, no.~1, pp. 173--196, 2019.

\bibitem{9625826}
D.~Dardari, N.~Decarli, A.~Guerra, and F.~Guidi, ``{LOS/NLOS} {N}ear-{F}ield {L}ocalization {W}ith a {L}arge {R}econfigurable {I}ntelligent {S}urface,'' \emph{IEEE Trans. Wireless Commun.}, vol.~21, no.~6, pp. 4282--4294, 2022.

\bibitem{9838395}
B.~Luo, M.~Dong, H.~Wu, Y.~Li, L.~Yang, X.~Chen, and B.~Bai, ``Reconfigurable {I}ntelligent {S}urface {A}ssisted {M}illimeter {W}ave {I}ndoor {L}ocalization {S}ystems,'' in \emph{2022 IEEE International Conference on Communications (ICC)}, 2022, pp. 4535--4540.

\end{thebibliography}
\end{document}